\documentclass{ws-procs975x65}
\usepackage{hyperref}
\begin{document}

\title{CAN THE CHAMELEON MECHANISM EXPLAIN COSMIC ACCELERATION WHILE SATISFYING SOLAR SYSTEM CONSTRAINTS ?}

\author{A. HEES$^{1,2,3,*}$ and A. F\"UZFA$^{1}$}

\address{$^1$ Namur Center for Complex System (naXys), University of Namur, Belgium\\
$^2$ Royal Observtory of Belgium (ROB), Avenue Circulaire 3, 1180 Bruxelles, Belgium\\
$^3$ LNE-SYRTE, Observatoire de Paris, CNRS, UPMC, 75014 Paris, France\\
$^*$E-mail: aurelien.hees@oma.be}

\begin{abstract}
	The chameleon mechanism appearing in massive tensor-scalar theory of gravity can effectively reduce the nonminimal coupling between the scalar field and matter. This mechanism is invoked to reconcile cosmological data requiring introduction of Dark Energy with small-scale stringent constraints on General Relativity. In this communication, we present constraints on this mechanism obtained by a cosmological analysis (based on Supernovae Ia data) and by a Solar System analysis (based on PPN formalism).
\end{abstract}

\keywords{Dark Energy; Alternative theories of gravity; Chameleon fields}

\bodymatter

\section{Introduction}
General Relativity (GR) passed all Solar System tests until now. However, cosmological observations are not directly compatible with GR and the standard model of elementary particles. A way of modifying gravity at large scale without doing any modification at Solar System scale is achieved by the so-called chameleon mechanism~\cite{khoury:2004fk,khoury:2004uq}. This mechanism appears in tensor-scalar theories of gravity with massive scalars. In this communication, we explore the sensitivity of chameleon mechanism by constraining its parameters by cosmic acceleration on cosmological scales and by PPN constraints on small scales~\cite{hees:2012kx}.
This combined analysis will shed new light on the question whether modified gravity can be safely invoked to solve cosmological problems without any contradiction on Solar System scales.                            

\section{Model}
The model considered here is the one proposed by Khoury and Weltman~\cite{khoury:2004fk,khoury:2004uq} i.e. a tensor-scalar theory of gravitation with a runaway potential $V(\phi)$ (with $c=1$):
\begin{equation}\label{action}
S=\int d^4x\sqrt{-g}\left[\frac{m^2_p}{16\pi}R-\frac{m^2_p}{2}\partial_\mu \phi \partial^\mu \phi - V(\phi)\right]+S_M(\Psi_m,A^2(\phi)g_{\mu\nu})
\end{equation}
where $R$ is the scalar curvature, $m_p$ is the Planck mass ($m_p^2=1/G$) and $A(\phi)$ is a coupling function here given by $A(\phi)=e^{k\phi}$ ($k$ beeing the coupling constant). The chosen potential is a Ratra-Peebles potential~\cite{peebles:1993fk} $V(\phi)=\frac{\Lambda^{4+\alpha}}{m_p^\alpha\phi^\alpha}$ parametrized by two constants $\alpha$ and $\Lambda$.

The action (\ref{action}) is expressed in the so-called {\bf Einstein frame}. This frame is useful for doing calculations but all the physical interpretations are easily done in the {\bf Jordan frame} where matter is minimally coupled to the metric $\tilde g_{\mu\nu}=A^2(\phi) g_{\mu\nu}$~\cite{damour:1992ys,damour:1993kx}. Quantities expressed in Jordan frame are noted with a tilde and observables can be computed as in GR within this frame~\cite{flanagan:2004fk}.

\section{Cosmological constraints}
Fields equations have been derived by introducing a Friemann-Lemaitre-Robertson-Walker Einstein-frame metric ($ds^2=-dt^2+a^2d\ell^2=a^2(-d\eta^2+d\ell^2)$ with $\eta$ the conformal time) and by replacing the Einstein frame matter density/pressure by the observable one (expressed in Jordan frame~\cite{damour:1992ys,damour:1993kx}: $\rho=A^4(\phi)\tilde \rho$, $p=A^4(\phi)\tilde p$).
The distance-luminosity relation
\begin{equation}
	\mu=25+5\log\left(\frac{\eta_0-\eta}{A(\phi)a}\right)	
\end{equation} 
(with $\eta$ expressed in \rm{MPc} and $\eta_0$ denoting the conformal time at actual epoch characterized by $A(\phi_0)a_0=1$) has been derived in both Einstein and Jordan frames to show their physical equivalence~\cite{hees:2012kx}.        

Each cosmological model is characterized by 4 parameters: the coupling constant $k$, the parameters entering in the expression of the potential $\alpha$ and $\Lambda$ and the actual observable matter density $\tilde\Omega_{m0}$. The value of the energy scale of the potential $\Lambda$ is optimized such that for a given value of $\alpha$, the input value of $\tilde \Omega_{m0}$ is retrieved. As shown in~\fref{figComb} (a), the value of $\Lambda$ depends mainly on the value of $\alpha$.

For the three other parameters, a likelihood analysis has been performed on Supernovae Ia measurements (with the UNION dataset~\cite{kowalski:2008zr}). The  68 and 95 \%  confidence regions have been derived and depend on the coupling constant $k$~\cite{hees:2012kx}.
\def\figsubcap#1{\par\noindent\centering\footnotesize(#1)}          
\begin{figure}[b]%
\begin{center}
\parbox{2.1in}{ \includegraphics[width=1.95in]{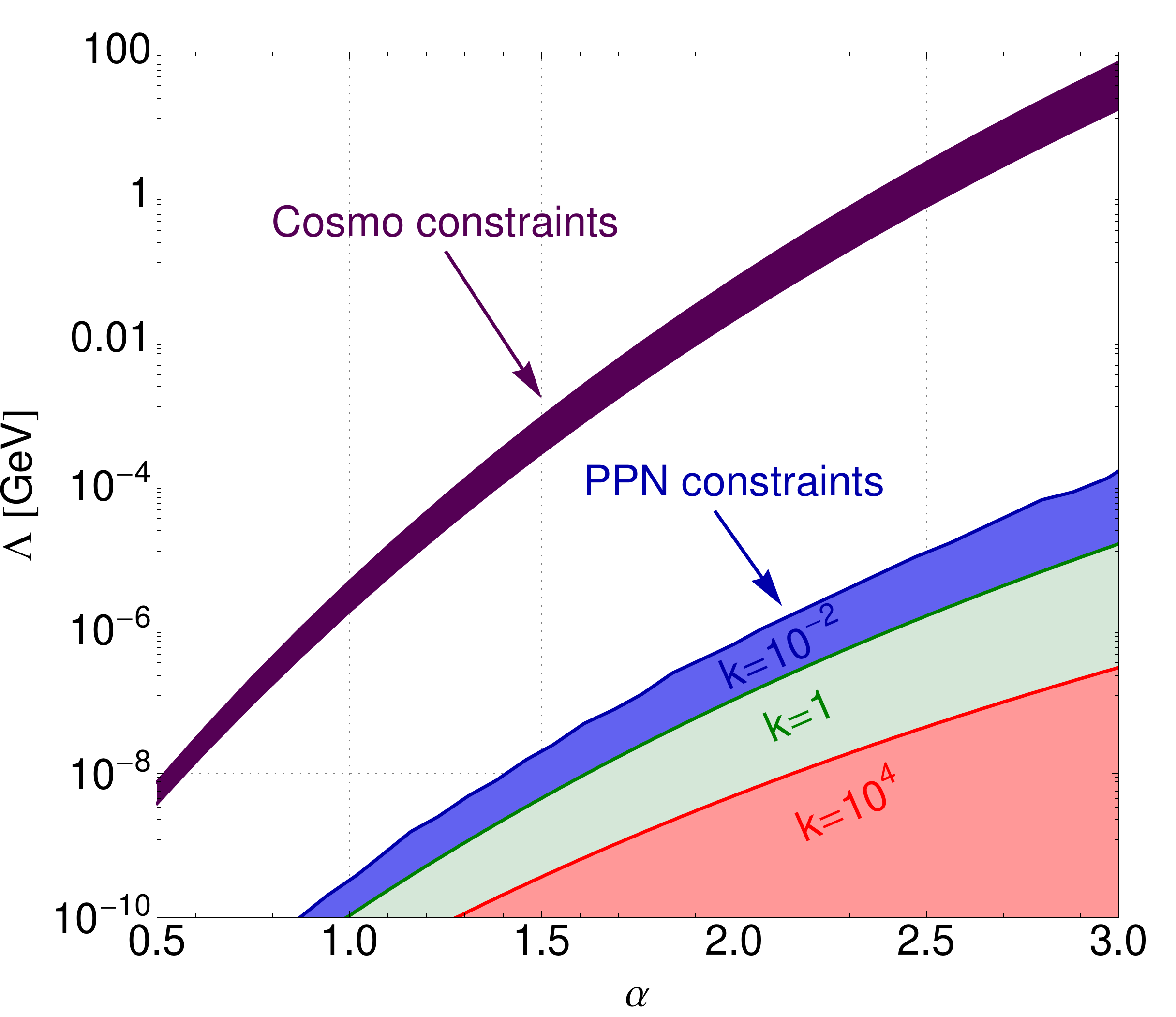}
\figsubcap{a}}
\hspace*{.3in}
\parbox{2.1in}{\includegraphics[width=1.8in]{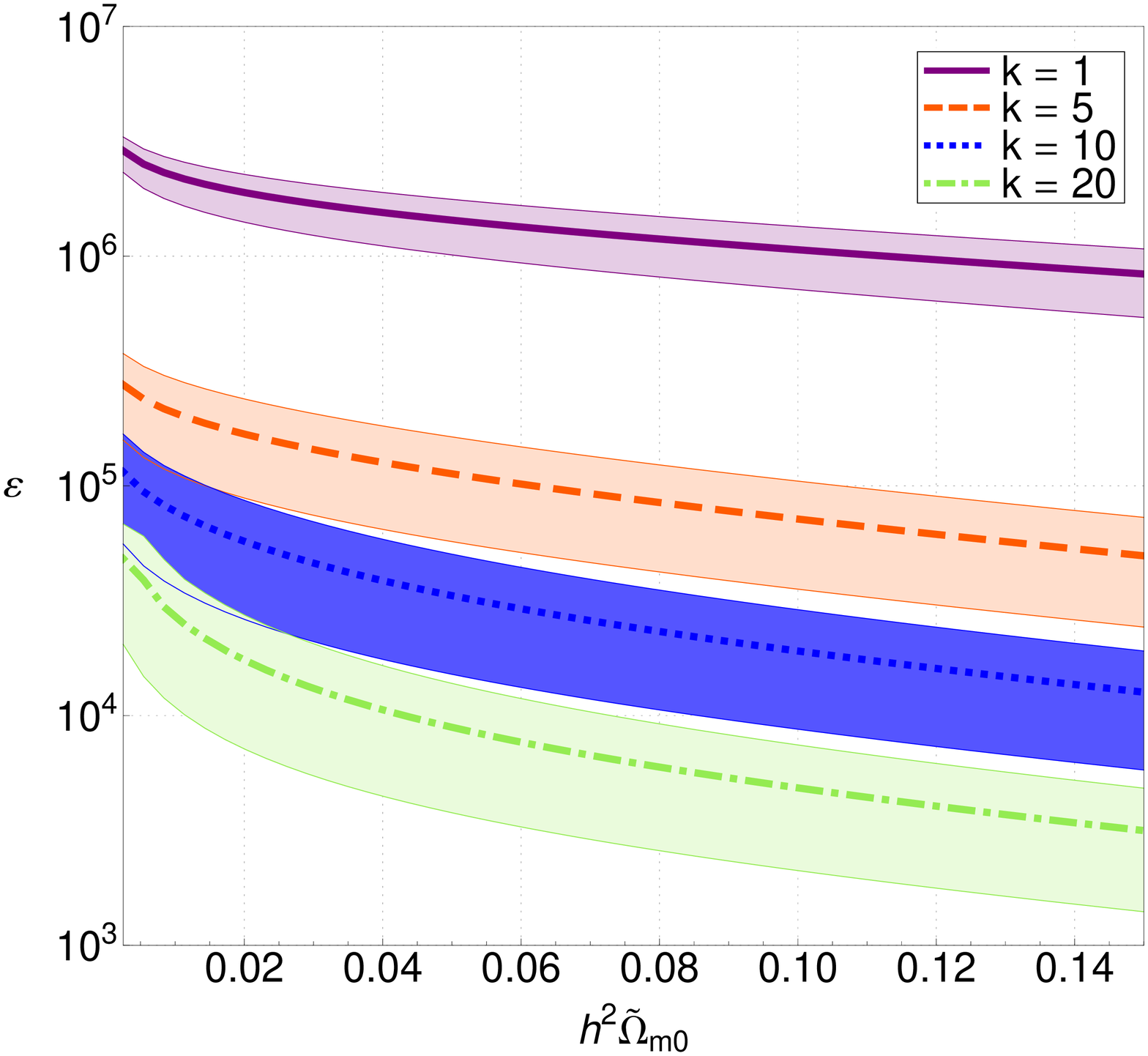} 
\figsubcap{b}}
\caption{(a) Representation of the cosmological constraint between $\Lambda$ and $\alpha$ and of the region allowed by the PPN constraint for different value of the coupling constant $k$.  \newline
(b) Representation of the thin-shell parameter $\epsilon$. The different lines represent different values of the coupling constant and the filled areas represent different values of $\alpha$ between 0.5 and 3.}
\label{figComb}
\end{center}
\end{figure}

\section{Solar System constraints}
The static spherical solution of the field equations deriving from action (\ref{action}) representing the Sun has been derived analytically by Khoury and Weltman~\cite{khoury:2004fk} and by Tamaki and Tsujikawa~\cite{tamaki:2008ve}. We confirm that the underlying hypothesis used in these papers are correct with a comparison with numerical simulations~\cite{hees:2012kx}. In the Solar System, the key parameter is the {\bf thin-shell parameter $\epsilon$}~\cite{khoury:2004fk,khoury:2004uq}. If $\epsilon>1$, the theory is equivalent to a Brans-Dicke theory (the potential does not play any role in the Solar System) and the PPN constraint~\cite{bertotti:2003uq} on $\gamma$ gives the traditional constraint $k^2<10^{-5}$. If $\epsilon << 1$, the potential plays a crucial role and the deviation from GR can be screened in the Solar System. In particular, the $\gamma$ PPN parameter depends explicitly on the parameters entering the potential ($\alpha$ and $\Lambda$) and the PPN constraint can be satisfied even for high coupling constant\cite{hees:2012kx}. Regions of parameters satisfying the PPN constraint are represented on \fref{figComb} (a).

\section{Conclusion}
The analysis at two different scales sketched in the previous sections shows the chameleon mechanism can not be invoked to explain cosmic acceleration while satisfying Solar System constraints. The analysis in the $(\alpha,\Lambda)$ plane (presented in \fref{figComb} (a)) shows there is no intersection between regions of parameters reproducing cosmological observations and satisfying PPN constraints. Another way to reach this conclusion is to consider the thin-shell parameter for all models within the 68~\% cosmological confidence region. As shown in \fref{figComb} (b), $\epsilon >> 1$ for models explaining SNe Ia data, which means the potential does not play any role in the Solar System and the theory is equivalent to Brans-Dicke\cite{hees:2012kx}. This conclusion is only valid for the Ratra-Peebles potential and the exponential coupling function considered here.

\section*{Acknowledgments}
A. Hees is supported by an FRS-FNRS Research Fellowship. Numerical simulations were made on the local computing resources (cluster URBM-SysDyn) at the University of Namur (FUNDP).                                                               

\bibliographystyle{ws-procs975x65}
\bibliography{../../../../byMe/biblio}

\end{document}